\documentclass[prl,twocolumn,showpacs,superscriptaddress]{revtex4-1}

\usepackage{graphicx}
\usepackage{dcolumn}

\usepackage{eucal}
\usepackage{epstopdf, wrapfig}
\usepackage[dvips]{epsfig}
\usepackage{amssymb}

\usepackage{amsmath,amsthm}
\usepackage{natbib}

\include{tex_options}

\begin{document}
\title{Glassy dynamics in three-dimensional embryonic tissues}
\author{Eva-Maria Sch\"otz}
\author{Marcos Lanio}
\author{Jared A. Talbot}
\affiliation{170 Carl Icahn Laboratory, Lewis-Sigler Institute, Princeton University, Princeton, NJ 08544, USA}
\author{M.~Lisa Manning}
\email{mmanning@syr.edu}
\affiliation{Department of Physics, Syracuse University, Syracuse, NY 13244, USA}

\date{\today}

\begin{abstract}
Many biological tissues are viscoelastic, behaving as elastic solids on short timescales and fluids on long timescales.  This collective mechanical behavior enables and helps to guide pattern formation and tissue layering.  Here we investigate the mechanical properties of three-dimensional tissue explants from zebrafish embryos by analyzing individual cell tracks and macroscopic mechanical response.  We find that the cell dynamics inside the tissue exhibit features of supercooled fluids, including subdiffusive trajectories and signatures of caging behavior.  We develop a minimal, three-parameter mechanical model for these dynamics, which we calibrate using only information about cell tracks.  This model generates predictions about the macroscopic bulk response of the tissue (with no fit parameters) that are verified experimentally, providing a strong validation of the model.  The best-fit model parameters indicate that although the tissue is fluid-like, it is close to a glass transition, suggesting that small changes to single-cell parameters could generate a significant change in the viscoelastic properties of the tissue. These results provide a robust framework for quantifying and modeling mechanically-driven pattern formation in tissues. 
\end{abstract}

\maketitle


\section{Introduction}
A quantitative description of the mechanical behavior of groups of cells in biological tissues is critical for an understanding of many fundamental biological processes, including embryogenesis~\cite{Schoetz2008, keller_forces_2008, Farge2003, Farhadifar, McMahon2008, sandersius2011,Weijer2009}, wound healing~\cite{Poujade2007, Schneider2010}, stem cell dynamics, regeneration~\cite{Soriano2009, Laird2008}, and tumorigenesis~\cite{Christiansen2006, Friedl2009, Tse2012, Hegedus2006}.  Previous work demonstrates that the macroscopic response of many tissues is viscoelastic~\cite{Schoetz2008}, where the tissue behaves as an elastic solid over short timescales and a viscous fluid over long timescales.  The {\em relaxation timescale}, a material parameter that is different for different tissue types, characterizes how much time it takes for a tissue to cross over from elastic to viscous behavior.

  Recently researchers have discovered that some two-dimensional tissues exhibit glassy behavior~\cite{Angelini, Kaes}.  In these confluent tissues, individual cells have difficulty moving past one another or exchanging neighbors, resulting in a ``frozen'' system with a macroscopic response that is solid-like even on long timescales. A wide variety of materials exhibit glass transitions, which generally occur when the individual agents that comprise the material (i.e. atoms, polymers, droplets, cells) approach high densities or low individual motilities (i.e. atoms at low temperature).  Glasses display universal features, including long-range spatial correlations in velocity fields and slowing down of individual agents~\cite{Debenedetti, Ediger, Abate}.  Identifying these features in tissues is useful because existing theoretical models for glasses can be adapted to predict how the properties of single cells generate a collective macroscopic response~\cite{Henkes, Berthier}.

   Because embryonic explants are fluid-like on long timescales, they are not glasses. However, embryonic tissues share several properties with another class of materials called super-cooled fluids -- they are tightly packed in disordered structures and display elastic behavior on short timescales and fluid behavior on long timescales. Supercooled fluids have a viscoelastic relaxation timescale that is controlled by their proximity to a glass transition~\cite{Richert, Debenedetti} and display glassy dynamics such as anomalous slowing of individual motions and “caging” behavior~\cite{Kob1997, Weeks2002}. To test whether the viscoelasticity of embryonic tissues is similarly controlled by proximity to a glass transition, we must first determine whether cell trajectories in embryonic tissues display signatures of glassy dynamics, and then we must develop a framework for determining how “close” the system is to a glass transition. Finally, we would like to use this information to make quantitative predictions about macroscopic tissue viscoelasticity.  

The remainder of this paper is organized as follows: first, we track individual cells in three-dimensional zebrafish explants and demonstrate that cell trajectories exhibit anomalous diffusion and caging behavior, which are both signatures of glassy dynamics.

Second, we develop a novel minimal model with three dimensionless parameters that can be used to quantify how close the material is to a glass transition.  The parameters have simple biophysical interpretations: a) the ratio between the adhesion and the cortical tension, b) the magnitude of  forces actively generated by cells exerting tension on their neighbors, and c) the typical timescale over which those forces act.  As a function of the three parameters, we find that the model does have a ``jamming'' or glass transition and demonstrate that the best-fit model parameters describe a supercooled fluid that is controlled by this nearby transition.

Third, we use this calibrated model with no fit parameters to make predictions about macroscopic viscoelastic response as measured by tissue compression and tissue fusion assays.  We demonstrate that our minimal model accurately predicts the viscoelastic relaxation timescale seen in both types of experiments, providing a strong validation of the model. The model also reproduces qualitative observations of surface properties, such as the existence of a tissue surface tension, but it does not correctly reproduce the magnitude of the tissue surface tension. This is consistent with previous work~\cite{Manning_PNAS, Manning_Science} suggesting steady-state surface properties are sensitive to the detailed shapes and tensions of individual cells, which are not included in our model.

  We conclude with a discussion of how these results can be extended to better understand more complicated biological processes.  Now that we have a good handle on the mechanical behavior in simple environments, we can systematically add additional degrees of freedom to the model to capture biochemical signaling near tissue boundaries, coupling to extracellular matrix, or chemotaxis that occurs in signaling gradients.

\section{Materials and Methods}

\subsection{Experimental methods}
\textbf{Explant preparation} Maternal Zygotic \textit{one-eyed pinhead} (MZ\textit{oep}) and 50-100pg of \textit{cyclops} mRNA-injected wildtype zebrafish aggregates, corresponding to ectoderm and mesendoderm, respectively, were generated and fluorescently labeled as previously described~\cite{EMSthesis, Schoetz2008, Manning_PNAS}.

\textbf{Explant fusion} Fusion experiments were carried out on a Zeiss Axiovert 200M microscope (Zeiss, G\"{o}ttingen, Germany), equipped with a SPOT camera (Diagnostic Instruments, Inc., MI) and Metamorph 4.6 (Molecular Devices, LLC, CA) software and on a Olympus DSU microscope (Tokyo, Japan), equipped with a Hamamatsu camera (Hamamatsu City, Japan) and Slidebook 5.1 software. The aspect ratio, AR, of fusing tissue aggregates was determined by finding the object's convex hull using a built-in MATLAB (MathWorks; Natick, MA) function, and extracting its major and minor axes (see SI). The aspect ratio as a function of time was plotted and fitted with an exponential$+$constant~\cite{Cliff2011}.

\textbf{Tissue surface tension measurements} The tissue surface tensiometer (TST) was constructed as previously described~\cite{Foty1994,Foty1996}, without the water circulation and with a digital CAHN electrobalance D-200 (Cerritos, CA) and the lower compression plate (LCP) connected to a Newport NewStep NSA12 actuator (Irvine, CA), allowing for computer-controlled motion through a custom LabVIEW program (National Instruments, Austin, TX). Aggregates were imaged using a Basler A601f camera (Ahrensburg, Germany) attached to a Leica S8APO stereo microscope (Wetzlar, Germany). The tensiometer was calibrated as described in the SI. Tissue surface tension and Young's modulus were calculated as described in~\cite{Mgharbel2009, EMSthesis}.

\textbf{2-photon imaging and cell tracking} For time-lapse \textit{in vivo} imaging, explants were embedded in 1$\%$ low-melting point agarose (Cat.nb 15517-022, Invitrogen) in E2-medium~\cite{Schoetz2008} to minimize motion. Imaging was carried out on a custom-built two-photon laser scanning microscope, constructed on an upright BX51 Olympus microscope (Tokyo, Japan). A tunable Ti:Sapphire pulse laser (Mira 900, Coherent,  100-fs pulses at 80 MHz) was used to excite the sample with $\sim$920nm pulses. The emitted light was collected simultaneously through an Olympus LUMPlan Fl/IR water-immersion objective (NA$=$0.8) and an oil-immersion condenser (NA$=$1.4) using GaAS photomultiplier tubes (Hamamatsu Photonics K.K; Hamamatsu City, Japan). MATLAB ScanImage software was used for image acquisition~\cite{12801419}.  Fluorescently labeled nuclei were identified in Z-stacks of 2-photon images, with one Z-stack captured every two minutes, and resolution of 0.86 microns/pixel in X and Y and 4 microns/pixel in Z.  The images were analyzed using a bandpass filter and 3D feature finding algorithms~\cite{Crocker1996}.  Features were binned according to position and volume and spurious features were identified as those with volumes less than 20 percent of average or located outside the spherical aggregate shape.  Features identified at each time point are linked into nuclei tracks according to a standard tracking algorithm~\cite{Crocker1996}, and automated tracks are compared to manual tracks to ensure accuracy. Because images for the ectoderm explants are slightly noisier, our tracking algorithm occasionally wrongly identifies noise as nuclei ``features'', but these mislabeled features have very short tracks (2-3 frames) and do not affect results at timescales longer than 5 mins.

\section{Results}

\subsection{Statistics of individual cell trajectories}

\begin{figure}[h!]
\centering \includegraphics[width=0.49\textwidth]{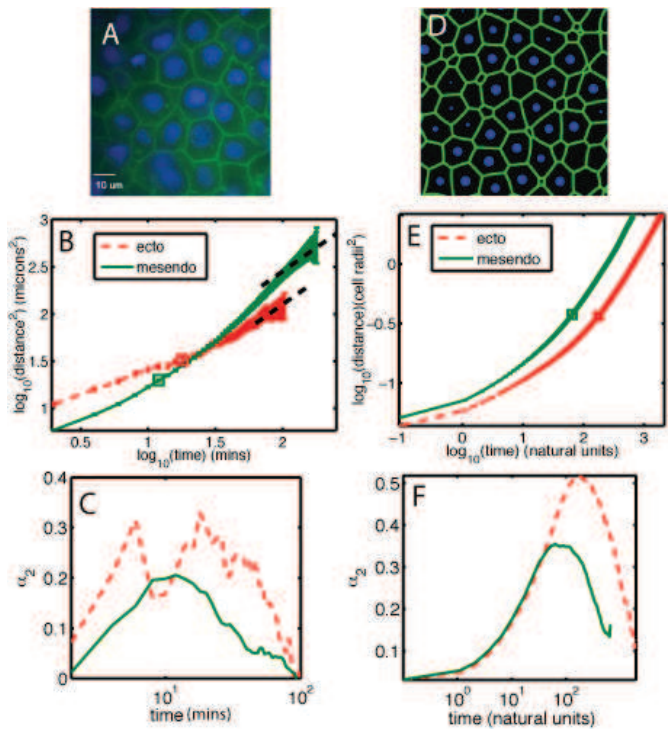}
\caption{\label{fig_structure} (A) Experimental tissue showing packing fraction unity. Cell membranes are labeled using Gap43GFP, cell nuclei using Hoechst. (B) Mean squared displacement (MSD) data for experimental ectoderm (red dashed) and mesendoderm (green solid) explants. Black dashed lines are slope 1, drawn to guide the eye. (C) Experimental non-gaussian parameter (described in text) achieve their maxima at a crossover timescale $t_c$. Secondary peak at very short timescales for ectoderm data is likely caused by misidentified features.  (D) Simulation of experimental tissue, interfaces generated using voronoi tessellation and surface evolver. (E) MSD data for best-fit simulation parameters shown in natural units. (F) Non-gaussian parameters calculated for best-fit simulation parameters.}
\end{figure}

We first analyze the structure and dynamics of cells in ectoderm and mesendoderm zebrafish explants. We reconstruct the three-dimensional static positions for a subset of the nuclei in the explant at each timepoint, and estimate cell shapes by taking a 3D voronoi tessellation~\cite{voronoi} of the nuclei positions. In both tissue types, the structure of the tissue is disordered; the cell nuclei are not arranged in a crystalline pattern, and the cell shapes are irregular polyhedra with roughly similar volumes.  A two dimensional slice through the tissue therefore appears as curved polygons with widely varying areas (Fig.~\ref{fig_structure}A).

A second observation is that the tissue is confluent, where there are no visible extracellular gaps in membrane-labeled images. One way to quantify a cellular structure is the dimensionless packing fraction $\phi$, which is the ratio of the sum of the volumes of all the individual cells compared to the total volume taken up by the aggregate.  For tissues with extracellular gaps the packing fraction is less than one, but for completely confluent tissues the packing fraction is unity. This value can be directly compared with results from simulations.

To non-dimensionalize other observables, we define the average effective radius $R$ of cells by calculating the average distance between nuclei in the middle of the aggregate, which is $15 \pm 2 \mu m$. Since the overlap between soft disordered spheres at packing fraction unity is approximately 15 \%~\cite{Ohern},  we find that twice the effective radius averages 17 $\mu m$ and the average effective radius is $ R = 8 \pm 1 \mu m$.

By combining the static three-dimensional positions of nuclei from different timepoints, we can track them over time and analyze their dynamics inside the tissue explants~\cite{Crocker1996}.  

   A standard metric for studying the motion of particles is the mean squared displacement (MSD), which is the square of the net distance an individual particle moves as a function of time, averaged over all particles. The motion of the nuclei is diffusive if the MSD scales linearly with time, and super (sub)-diffusive if the MSD increases with the time to a power greater (less) than one.   For diffusive tissues in three dimensions, the diffusion constant $D$ is one-sixth the long-time limit of the ratio between the MSD and time.

Figure~\ref{fig_structure}B shows the log of the MSD as a function of log of the time for ectoderm and mesendoderm tissues. We find that $D = 0.22 \pm 0.05\mu m^2$/min for ectoderm and $D= 0.60 \pm 0.05 \mu m^2$/min for mesendoderm.   The slope of this plot,  which is a second dimensionless observable $\alpha$, is nearly unity at long times for both tissues indicating that the nuclei movement is diffusive.  Furthermore, we observe that both tissues exhibit a subdiffusive regime at short timescales where the slope is significantly smaller than unity on a log-log plot.

   While there are many instances in biology where a material exhibits sub-diffusive behavior, these MSD curves are reminiscent of those seen in supercooled colloids that are just above the glass transition temperature~\cite{Weeks2002}. In those materials, individual colloidal particles are trapped in a ``cage'' of neighbors, and they must wait for a rare, high-energy temperature fluctuation to escape that cage and continue diffusing. Given that the tissues are at packing fraction unity, it takes a significant amount of mechanical energy for one cell to move past another, and therefore the subdiffusive regime in tissues could be generated by a similar mechanism.

 To test this hypothesis,  we analyzed images of the real-space trajectories of individual nuclei inside ectoderm explants.  We find several trajectories that exhibit signatures of caging behavior  -- long periods where cell displacements are small compared to the cell radii punctuated by a period of directed travel that traverses roughly a cell diameter. (Example trajectories are shown in Fig S5 in the SI).  To systematically investigate these caging effects, we calculate the average non-gaussian parameter for all cell trajectories~\cite{Kob1997}, which quantifies the directedness of cell trajectories as a function of timescale.  We expect cell displacements to be random and roughly gaussian-distributed at short timescales when they are caged by their neighbors, and then again at very long timescales after they have exchanged many neighbors.  However, at intermediate timescales when they break out of their cages, we expect the non-gaussian parameter to be large because the trajectories are more directed and less random.   Fig.~\ref{fig_structure}(C) shows the non-gaussian parameter $\alpha_2$ as a function of timescale, and demonstrates that there is a peak in the non-gaussian parameter at roughly the same timescale as the crossover from sub-diffusive to diffusive behavior in the MSD trajectories: $t_c \sim 20 $ mins for ectoderm and $t_c \sim 10$ mins for mesendoderm tissues. This is consistent with our hypothesis that cell trajectories are supercooled or caged, and allows us to define a third dimensionless \emph{cross-over time} observable $\tau^* = D t_c/(R^2)$.   For the ectoderm aggregates, a spurious peak occurs at very short timescales (3-5 minutes) due to mislabeled features in our tracking algorithm, but as discussed above this does not affect the results for longer tracks. 

\subsection{A Minimal model for cell dynamics}

This model attempts to determine the minimal ingredients necessary to explain the macroscopic bulk viscoelastic response of simple embryonic tissues. Instead of allowing many degrees of freedom per cell, corresponding to the viscoelasticity and activity of the actin-myosin cytoskeleton, we allow one degree of freedom per cell, the center of mass (COM), and introduce several types of {\em interactions} between cells to capture single cell viscoelasticity, adhesion, and active force generation.

We hypothesize that the emergent mechanical behavior of a large group of cells does not depend on detailed cell shapes and activities, but instead on a small set of variables that govern the rates at which cells can squeeze past one another in this tightly packed, disordered structure. Therefore, we focus on determining the correct length and time scales for typical cell-cell interactions, and later verify that the exact forms for these interactions are not important for determining the emergent behavior.  We identify four general classes of interactions that occur between cells: resistance to shape changes and adhesion (captured by an interaction term $\vec{F}^{int}$), damping ($\vec{F}^{damp}$), and active cell motility ($\vec{F}^{a}$).  A detailed description of each of these terms as well as their mathematical representation is given in the SI.  The most important difference between this model and others in the literature is that the active forcing term is not random white noise~\cite{Palsson, Ranft, Chate, Henkes}.  Instead, it is structured to be more biologically realistic; it enforces that cells can only move by exerting tension on adhesive contacts with other cells, and incorporates a timescale $p_t$ that characterizes how much time a cell typically spends moving in the direction of one adhesive contact before switching directions to move towards a different adhesive contact.

Assuming that the cell dynamics are overdamped the equation of motion for each cell $i$ is:
\begin{equation}
\label{Fsum}
\vec{0} = \vec{F}^{damp}_{i} + \sum_{<ij>} \vec{F}^{int}_{ij} + \sum_{<ij>} \vec{F}^{a}_{ij}.
\end{equation}

We nondimensionalize the equations with units of length equal to the average cell radius $R$, units of force equal to the product $K \, R$, where $K$ is an effective spring constant that characterizes the cortical tension, and time $\tau = b/K$ where $b$ is a damping coefficient.  The equation of motion for the position of the center of mass $r_{i}$ for cell $i$:
\begin{equation}
\label{dim_model}
\frac{d {\widetilde r_{i}}}{d\widetilde{t}} = - \sum_{j} \left[\left( \widetilde{\delta}_{ij}  - \widetilde{\Gamma} \right) \widehat{r}_{ij} + \widetilde{\sigma} \; \xi \; \widehat{a}_{ij} \right],
\end{equation}
where $\widehat{r}_{ij}$ is the unit vector connecting the two cell centers, and $ \widetilde{\delta}_{ij}$ is the cell overlap (how close two cells centers are compared to their radii). $\widehat{a}_{ij}$ is a unit vector in the direction of the active force calculated as described in the SI, and $\xi$ is a unit variance chi-distributed random variable with $k=3$ and persistence time $\widetilde{p_t}$. There are three dimensionless parameters $\widetilde{\Gamma} = \frac{2 \pi \gamma }{K}$, which is the ratio between the adhesion energy and cortical tension, $\widetilde{\sigma} = \frac{\sigma}{K R}$, which is the ratio of the magnitude of the active forces to the cortical tension, and  $\widetilde{p_t} = p_t \frac{K}{b}$ which characterizes the persistence time for active forces.  Therefore we must identify three dimensionless observables in the simulations and experiments to calibrate the model.

We first perform a simulation of a rounded droplet that mimics the experiments in which we studied individual cell tracks. We integrate the equations of motion (Eq.~\ref{dim_model}) for slightly polydisperse cells from the droplet initial conditions with non-periodic boundary conditions. Simulation initialization is described in the SI text. The tissues in the simulations reach a steady state droplet volume after approximately 10-20 natural time units, and then we continue to run the simulations for approximately 500 natural time units.

 Figure~\ref{fig_structure}D is a reconstructed image of a 2D slice through the three dimensional tissue simulation.  The center of mass is denoted by a sphere of radius $0.5 R$ and the green lines (generated using the program Surface Evolver in 2D~\cite{Brakke}) minimize the surface tension between cells.  In the best-fit parameter regime, the simulations exhibit a disordered structure, as confirmed by an analysis of the pair correlation function (Supplemental Fig. S4).

\begin{figure}[h!]
\centering \includegraphics[width=0.4\textwidth]{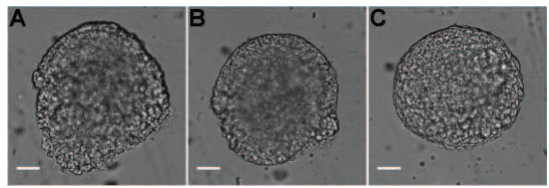}
\caption{\label{rounding} Brightfield still images showing a piece of zebrafish embryonic tissue round up. Scale bar is 30~$\mu$m. Supplemental movie S1 shows the entire sequence.}
\end{figure}

The simulated aggregates also show liquid-like dynamical behavior similar to that seen in the experimental cell aggregates. One of these behaviors is the rounding up of tissue fragments into a spherical shape~\cite{EMSthesis, Schoetz2008} as shown in Figure~\ref{rounding} and Supplemental Movie S1, which suggests that surface tension governs the final shape of aggregates.
 We observe that the qualitative behavior of the MSD and the nongaussian parameter in the simulations is similar to those seen in the experiments, as shown in Figure~\ref{fig_structure} (E,F). In addition, all cells remained part of a single connected cluster throughout the simulation, which is also generally seen in healthy experimental explants.
\begin{figure}[h!]
\centering \includegraphics[width=0.35\textwidth]{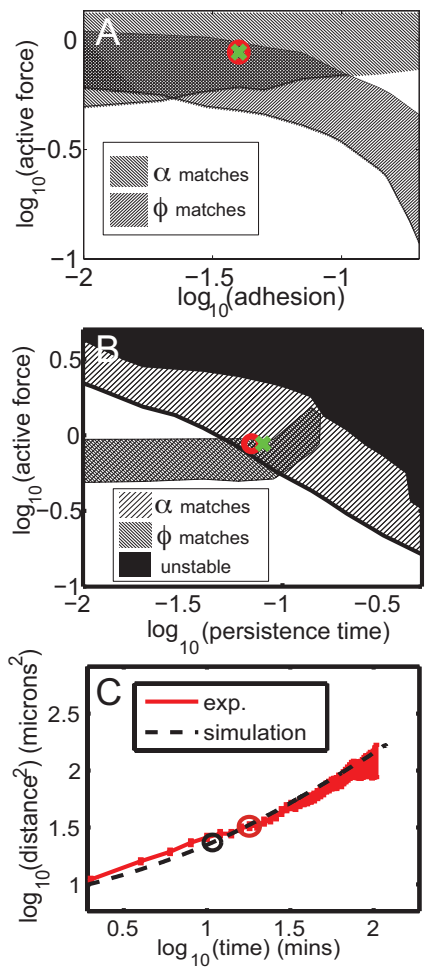}
\caption{\label{fig_phase} {\bf Calibrating the model} (A) Simulation phase diagram for active force and adhesion parameters. Red circle and green cross show best fit parameters for ectoderm and mesendoderm, respectively. (B) Phase diagram for active force and persistence time model parameters. (C) Comparison of MSD for ectoderm explants compared to best fit parameter from simulations.}
\end{figure}

Now that we have shown that the model is qualitatively similar in structure and dynamics to the experimental data, we calibrate it by varying the three dimensionless parameters ($\widetilde{\Gamma}$, $\widetilde{\sigma} $, $\widetilde{p_t} $) and identifying the best match with three dimensionless observables:  the product of the diffusion constant and the crossover timescale $\tau^* = D t_c /R^2$,  the scaling of the power law relationship between the MSD and time $\alpha$, and the packing fraction $\phi$.   Figure~\ref{fig_phase} (A, B) illustrate the results of hundreds of simulations with varying model parameters. Hatched regions correspond to parameter ranges where the simulation matches the experimentally observed $\phi$ and $\alpha$, respectively, while the red circle and green cross pinpoint the exact parameter values which match $\tau^*$ for the ectoderm and mesendoderm, respectively. Details are described in the SI. Figure~\ref{fig_phase}(C) demonstrates that the functional form of the MSD for the best-fit ectoderm simulation is very similar to that for the experimental data. Table~\ref{table_best} summarizes the best fit parameters and conversion factors from our simulation model to both tissue types.

\begin{table}[h!]
\begin{tabular}{  l | c  }


  parameter & value \\
 \hline
  $\widetilde{\Gamma}$ &  0.04 \\
  $\widetilde{\sigma}$ & 0.88 \\
  $\widetilde{p_t}$ & 0.07 (ecto)  0.08 (meso) \\
  $\tau$ & 4 s \\
  $R$ & 8 $\mu m$
\end{tabular}
\caption{\label{table_best} Tissue simulation parameters}
\end{table}

We also calculate how sensitive the observables (i.e. the diffusion constant and packing fraction) are to changes in the model parameters (See SI Table~1) near the best-fit parameter values. The diffusion constant is very sensitive to changes in all three parameters, and most sensitive to changes in  $\widetilde{\sigma}$, which means that the model parameters are strongly constrained by the diffusion data. For example, changing the active force magnitude by 15\% changes the diffusion constant by 100 \%.  In contrast, the packing fraction $\phi$ is less sensitive to changes in model parameters. 

In addition, the model exhibits a jamming or glassy phase transition when the active forcing magnitude and persistence time are smaller than the best-fit values for ectoderm and mesendoderm. As discussed in the SI, we define the glass transition as the point at which the diffusion constant $D > 1 \times 10^{-4}$ in natural units, because we find that this coincides with the onset of dynamical arrest. Although the best-fit model parameters are in a regime that is not jammed, the transition is nearby.  For example, the model predicts that reducing either the active forcing magnitude or the persistence time by 20 \% would result in a glassy tissue where cells can not migrate.  This suggests that the viscoelasticity observed in these tissue might be controlled by the nearby glass transition.

\subsection{Predictions for Macroscopic tissue response}

  As a test of the predictive powers of this model,  we keep the model parameters fixed at the values in Table~\ref{table_best}, and simulate the response of ectoderm tissues to large-scale mechanical perturbations such as compression and fusion with no adjustable parameters, and find qualitatively similar behavior. We then quantitatively compare the emergent mechanical responses and timescales to those observed in the experimental data, finding reasonable agreement for bulk properties, but disagreement for surface properties.

The first set of simulations for a quantitative comparison are tissue surface tension (TST) parallel plate compression tests, where we seek to replicate the experiment in which a cellular aggregate is compressed between two parallel plates.  In our model, the walls are represented by a 2D triangular crystalline array of particles lying in a plane, as discussed in the SI.   Because we represent the wall with a single layer of particles, the necessarily stiff interaction potential makes the wall artificially sensitive to small changes in the positions of cells.  Therefore while the average value of the force on the wall is physically meaningful, the fluctuations in the forces on the wall are larger than those seen in the experiments.

\begin{figure}[h!]
\centering \includegraphics[width=0.49\textwidth]{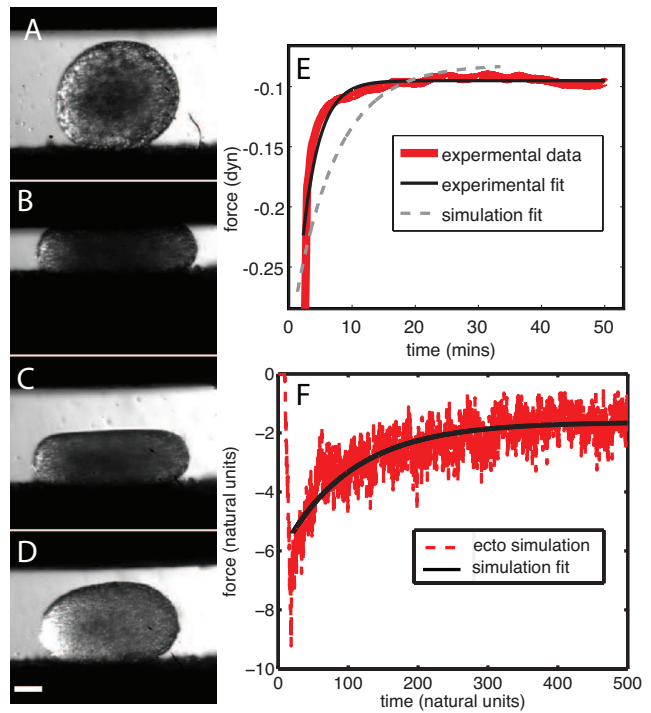}
\caption{\label{fig_TST} (A-D) Image series of an ectoderm explant compression in the TST: (A) Precompression - (B) compressed - (C) right after release - (D) $>$10min after compression. Scale bar is 100$\mu$m. (E) TST Experimental ectoderm force balance curve (thick red line) and fit (thin black line). For comparison, the normalized ectoderm simulation fit is shown by a gray dashed line. (F) Simulation force balance curves for ectoderm (red dashed line). Smooth black line is a fit to ectoderm data. }
\end{figure}

In both simulations and experiments we measure the net force exerted on the explant by the upper compression plate.  Images of an explant in a typical compression experiment are shown in Figure~\ref{fig_TST}(A-D). All of the experimental TST data demonstrates the same qualitative response shown in Figure~\ref{fig_TST} (E); as the plates are quickly brought together, there is a sharp decrease in the net force. This indicates a large force downward on the aggregate generated by the short-time elastic response of the aggregate. After the initial response, we observe a slow relaxation towards a non-zero equilibrium force as the cells rearrange and relax stress, just as a molecular fluid does.  The non-zero force at long times is generated by the effective surface tension of the tissue. Previous work has established that the extracted surface tension does not depend on amount of deformation, demonstrating that this is a surface effect and not a bulk effect~\cite{EMSthesis}. We fit the relaxation process to a constant$+$exponential. The exponential relaxation time of (4.5~$\pm$~0.8) min (mean $\pm$ SE; n$=$7) for ectoderm explants is a robust material property that describes the viscous tissue rheology.
The tissue surface tension $\gamma$ is obtained from the TST data using a modified Laplace's law which is more robust to fitting procedures for experimental data~\cite{Mgharbel2009}:
\begin{equation}
\label{gamma_TST}
\frac{F_{eq}}{\pi R_1^2} = \left( \frac{1}{R_2} - \frac{1}{R_1} \right) \gamma,
\end{equation}
where $F_{eq}$ is the steady-state force in the equatorial plane (of radius $R_1$) at long times, and $R_2$ is the local radius of curvature at the equator, extracted by fitting circles to brightfield images of the aggregate edge.  Finally we can also calculate the effective Young's modulus $Y$ for ectoderm aggregates for small compressions by assuming that the initial response of the tissue is elastic and applying a Hertzian model for the spherical object~\cite{EMSthesis}. We find that the Young's modulus for ectoderm is $44 \pm 11$ Pa ($n=5$), in agreement with previously published results of $Y = 48 \pm 9$ Pa~\cite{EMSthesis}.

The force-time response for a parallel plate compression of simulated ectoderm aggregates are shown in Figure~\ref{fig_TST}F. A first observation is that the mechanical response is qualitatively identical, capturing the features related to elastic, viscous, and surface tension effects.  In analogy to the experimental data, we fit the simulated data to a single exponential plus a constant, and find a relaxation time of 98~$\pm$~14 natural time units, which corresponds to (6.7~$\pm$~1.0) mins. The fact that with no fit parameters the simulated ectoderm has an emergent relaxation time which is similar to that for the experimental tissue is a strong validation of the predictive powers of the model.

To analyze whether the surface tension in the simulations matches experiments, we need an independent estimate of the force scale. If we assume that the Young's modulus for a single cell is the same as the Young's modulus for the entire aggregate (which is reasonable since the tissue is confluent) then $Y \sim 40-50 Pa$.  This means that the natural force units in our simulation should be approximately $\widetilde{F} \sim Y R^2 \sim 3 \times 10^{-4}$ dyn. An alternative path to the same result is to note that pipette aspiration experiments indicate that a typical effective cortical tension for tissues is $\gamma_c = 1$ dyn/cm~\cite{Evans}, which corresponds to a natural force unit for our simulation model of $\widetilde{F} \sim \gamma_c R \sim 8 \times 10^{-4}$~\cite{Durian}.  For the remainder of this paper, we use the value derived from TST data: $\widetilde{F} = 3 \times 10^{-4}$ dyn.

  Because the observables which are most robust to fits in the simulations are different from those in the experiments, we fit to a different form of Laplace's law that involves the radius of contact between the plate and the aggregate, $R_3$, as discussed in the SI. We find that $F_{wall} = 1.45$, $R_1=8.9, R_2=10.1$, and $R_3 =5.3$ in simulation units, which leads to a calculated value for the surface tension of $\gamma \sim 0.034$ dyn/cm.  This is about twenty-five times smaller than that seen in experiments, where $\gamma = 0.8 \pm 0.2 $ dyn/cm ($n=6$), and indicates that our model is not quantitatively capturing the surface tension effects in real tissues.

In hindsight this is perhaps not surprising, as we have shown in previous work that the surface tension depends sensitively on individual cell shapes at the surface of the cell aggregate, as well as temporary mechanical polarization of those cells~\cite{Manning_Science, Manning_PNAS}.  Since our simulation explicitly disregards cell shapes and polarizations, it does not quantitatively capture behaviors governed by surface tension, although it does capture qualitative features such as rounding up of aggregates. In the discussion, we will address how our model might be augmented to better capture surface tension effects.

A second test for the predictive powers of the model are tissue fusion experiments. For tissue fusion simulations, we initialize the system with two droplets of 1000 cells each and position them so that the average radii of the two droplets overlap by half of a single cell radius.  We then evolve all of the centers of mass according to Eq.~\ref{dim_model}.  

\begin{figure}[h!]
\centering \includegraphics[width=0.45\textwidth]{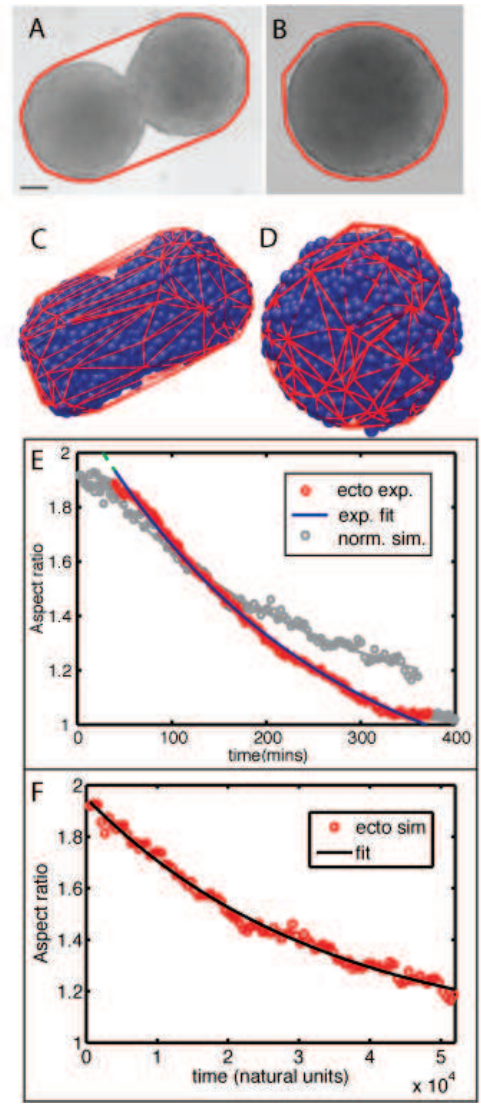}
\caption{\label{fig_fusion} (A,B) Images from the beginning and end of experimental aggregate fusion. Red line indicates convex hull. Scale bar is 100um. (C,D) Images from beginning and end of simulation of aggregate fusion. (E) Experimental data for aspect ratio as a function of time (red dots), and fit (blue line).  Gray circles and line correspond to simulation data that has been normalized to account for different aggregate size and surface tension as measured by the TST simulations.  (F) Simulation data for aspect ratio as a function of time (red dots) and fit (solid line).}
\end{figure}

 Figure~\ref{fig_fusion}(A,B) are snapshots from an experiment demonstrating that two rounded ectoderm explants join together to form a single rounded tissue (See Supplemental Movie S2).  To quantify this behavior, we first identify the convex hull of the two-dimensional images of the aggregate using standard image analysis techniques (see SI and Supplemental Movie S3), and the resulting hull is illustrated by red lines in Figure~\ref{fig_fusion}(A,B). Using a method similar to Ref.~\cite{Cliff2011}, we define the aspect ratio to be the ratio between major and minor axis, and Fig~\ref{fig_fusion}E illustrates that it varies from approximately two at the beginning of a fusion experiment to approximately unity when the fusion has completed for the example shown in Figure~\ref{fig_fusion}(A,B).  We fit the decay of the aspect ratio and find a characteristic decay timescale of $(207 \pm 23)~min$ (mean$\pm$SE; n$=$6) for ectoderm explants.

We also perform the same test on simulated ectoderm explants. Figure~\ref{fig_fusion}(C,D) are snapshots from the three-dimensional simulations, with the cells denoted by blue spheres and the red mesh denoting the convex hull. A full time sequence is available in Supplemental Movie S4. We again calculate the aspect ratio for each timepoint using the same definitions as for the experiments (scaled from 2D to 3D).  The resulting aspect ratio evolution, which is qualitatively very similar to the experimental observation, is shown in Figure~\ref{fig_fusion}F. The aspect ratio does not quite asymptote to unity because of the difference in the way the major and minor chords are calculated in 2D vs. 3D.  We fit this time evolution to the same exponential$+$constant function, and find a decay time of $33,500$ in natural units, or  $2,250$ mins.  This is more than an order of magnitude larger than the experimental observation, but there are two important differences between the simulated aggregate and the experimental aggregate. First, the experimental aggregate has about eight times as many cells as the simulated aggregate; for consistency we used a simulated droplet of the same size as those in the simulations in Figure~\ref{fig_phase} (A,B). Second, the surface tension of the simulated aggregate as measured by the TST simulation is about twenty times less than that for the experimental aggregate.  
In 1939 Young~\cite{Youngellipsoid} derived an analytic expression for the change in the aspect ratio ($r_a$) of an ellipsoidal droplet as it approaches a spherical droplet under the influence of surface tension:

\begin{equation}
\frac{d r_a}{dt} = \frac{1}{V^{1/3}}\frac{\gamma}{\eta} f(r_a),
\end{equation}
where $\gamma$ is the surface tension, $\eta$ is the viscosity, $V$ is the volume, and $f(r_a)$ is a complicated function that depends on whether the ellipsoid is oblate or prolate.  This equation indicates that we can account for the known differences between the simulated aggregate and the experimental aggregate by multiplying the time scale by $R_{exp}/R_{sim} \sim 2$ and dividing the timescale by the fit from the TST data $\gamma_{exp} / \gamma_{sim} \sim 23$.  The resulting curve for the ``effective'' aspect ratio has a time decay constant of $195$ mins, which compares remarkably well with the experimental timescale of $207$ min, given there are no fit parameters.  We also plot the effective simulated aspect ratio in light gray in Figure~\ref{fig_fusion}E, illustrating the similarity between the two.  Therefore, the simulated tissue fusion data is consistent with the experimental fusion data under the assumption that the simulated aggregate has the same bulk properties but a reduced surface tension.

\section{Discussion}

We have shown that a minimal model for cell interactions with three dimensionless parameters predicts several types of emergent, collective mechanical responses in embryonic tissues.  It disregards a significant amount of information about cell shapes and small-scale details of the mechanical interactions between cells. It is simple enough that it can be calibrated from experimental data using only single-cell trajectories and static nuclei imaging. Despite this simplicity it captures complex features of the experimental data, such as caging behavior and crossover timescales. When we additionally calibrate the force scale using TST data, the model can also predict the fusion behavior of two explants.

Our model indicates that macroscopic tissue behavior is most sensitive to the two model parameters that describe forces actively generated by cells.  Tissues are solid-like when cells generate weak forces or change their direction rapidly, and liquid-like when cells generate larger forces or change their direction slowly. In addition, tissues become more solid-like when the ratio between the adhesion and the cortical tension is high. Small changes to the model parameters lead to large changes in tissue behavior near the transition point, and the observation of caging and subdiffusive dynamics in embryonic tissues suggests that they are close to this transition.  Therefore, we anticipate that this model will be useful for making predictions about the macroscopic behavior of tissues or colonies composed of cells with motility defects. It should also be useful for predicting which tissues are fluid-like and therefore governed by tissue surface tension, and which tissues are solid-like and governed by elasticity.

If the specific details of single-cell mechanics are important for the large-scale mechanics of the tissue, we would expect our model to fail. The fact that the model succeeds does not mean that single-cell mechanics are not important for bulk properties, but suggests that details such as cell motility, directionality, elastic modulus, etc., can be successfully coarse-grained into a small number of properly chosen parameters.

Furthermore, we show that the motion of cells past one another, which must be generated by cells actively exerting tension on contacts with other cells, is best modeled by a special type of structured noise (both multiplicative and colored), instead of positional or angular white noise~\cite{Palsson, Ranft, Chate, Henkes}. This choice for the noise is motivated both by experimental observations at the single-cell scale and the fact that our simple model cannot reproduce the macroscopic observations without it. Why does such a simple model work so well?  As we and others~\cite{Angelini, Brochard} have noted, active tissues display glassy dynamics: the motion of individual cells is constrained because they are surrounded by tightly packed neighboring cells that impede their progress. It has been shown that the dynamics in non-biological glassy or jammed materials display universality: dynamical features do not depend on the details of the interactions~\cite{Haxton2011,Schmiedeberg2011}.  Our work suggests that models with only a few parameters could be adequate for describing active tissues near this glass transition.

An important observation about our experimental zebrafish explants is that the tissue boundary remains remarkably coherent during all mechanical perturbations, including hanging drop, TST compression, and tissue fusion experiments. No cells exit the aggregate, even though cells move over large distances on the inside and together behave collectively as a fluid. This remarkable property is not observed in typical non-biological materials with short-ranged interactions; if particles behave as a liquid inside the droplet, they also necessarily leave the droplet and generate a steady state vapor pressure in a closed system.  We note that ``vapor pressures'' also occur in many models for biological tissues.  For example the one introduced by Ranft et al~\cite{Ranft} does possess a regime where the bulk is fluid-like, but unless the model also has artificially long-ranged interactions, it will generate a significant vapor pressure in the fluid regime.
In contrast, the model introduced here has a self-generated boundary that reproduces the experimentally observed absence of a vapor pressure, because the direction of the noise acting on one cell depends on the location of the cell's neighbors. A cell does not ``push off'' another cell, but instead is biased to move in a direction where it can make new contacts. 

 Although there is little or no vapor pressure in the 3D system, recent experiments~\cite{Brochard} have shown that cell aggregates can ``wet'' adhesive substrates and generate a two-dimensional vapor pressure on the surface.  By replacing the non-adhesive walls in our TST experiments with adhesive passive particles, we would be able to model those behaviors. We expect that our model would very naturally recapitulate the presence of a two-dimensional surface vapor pressure and lack of a three-dimensional vapor pressure.

 Another interesting direction is to investigate modifications to our model that could help explain the much larger surface tensions seen in experiments compared to simulations.  We believe our model fails to correctly capture the magnitude of tissue surface tension because it likely depends on cell shapes and a strong feedback between adhesion and cortical tension that occurs at tissue interfaces, as discussed in other work~\cite{Manning_PNAS}.  We could augment our model to account for this by replacing the isotropic interaction given by SI Eq.~2, by a non-isotropic interaction for surface cells.

While the model was inspired by and calibrated using data from embryonic zebrafish explants, we expect that variations of this model will be applicable to a broad range of biological tissues. Because it is simple, it can be effectively calibrated for different tissue types. It bridges microscopic mechanical information (cell elastic modulus, cell surface tension, adhesion, rate of protrusions) and macroscopic tissue mechanics (tissue viscosity, tissue surface tension, tissue elasticity).  To expand the model to capture more complicated tissues than the ones studied here, one could group several spheres together to allow shape changes and cell divisions. Another possibility is to allow for higher order interactions between multiple cells; for cells which are very soft, the surface area in contact and the elastic interaction do not depend only on the two-body overlap, but also on the direction of the contact and on the locations of neighboring cells.  A cluster expansion borrowed from statistical physics might be able to address such effects.

\section{Conclusions}

We have studied the trajectories of individual cells in embryonic explants and find that although the tissue is fluid-like on long timescales, it displays features of glassy or supercooled dynamics, including subdiffusive mean squared displacements on short timescales and caging behavior.  This suggests that the tissue might act as a viscoelastic material where the viscoelastic response is governed by a nearby glass transition.

To explore this hypothesis, we have developed a three-parameter model that makes predictions for the emergent properties of simple embryonic tissues. Two key observations that we incorporate into our model are a) that cells are biased to move in the direction of their neighbors, and b) that this motion occurs over a time window which is not infinitesimally small compared to other timescales in the tissue. The model exhibits a glass transition between solid-like behavior and dynamical arrest (when active forces are small and occur over short time windows), and liquid-like behavior (when active forces are large and persistent).

  We calibrated the model using only the tracking and structural data from experiments on zebrafish embryonic explants.  The best-fit parameters suggest that embryonic tissues are liquid-like but close to the glass transition, and therefore our model can be used to help explain how small single-cell motility defects might lead to large differences in tissue response during development.

We verify that the calibrated model makes accurate quantitative and qualitative predictions about the macroscopic tissue response in parallel plate compressions and tissue fusion assays, including a coherent self-generated boundary.  Because this simple model can explain the bulk tissue response, the exact details about individual cell shapes and mechanics are not critical for this experimental system; instead, we can effectively coarse-grain those details into a few parameters that can easily be extracted from experiments. The fact that the model fails to explain surface properties suggests that cell shapes and mechanical polarizations play an important role in those processes.

From a practical standpoint, this model can easily be expanded to make predictions about other tissues in development and disease.  From an intellectual standpoint, it explains how disordered tissue structures and high cell densities generate a viscoelastic response. It also provides an explanation for the timescale for the crossover from elastic to viscous behavior as a ``caging'' timescale.  Because it is simple and yet different from existing models for active matter, it provides a framework for thinking about the types of phase transitions (such as jamming or flocking) that are possible in biological tissues.

\section{Acknowledgements}The authors thanks T.~Bacarian for initial cell tracking, S.~Thiberge and the Lewis-Sigler Institute imaging core facility for 2-photon usage, and J.~Shaevitz and S. Henkes for comments on the manuscript. E.-M.~S. is funded by the Lewis-Sigler Fellowship and the Burroughs Wellcome Fund CASI award. M. L. Manning acknowledges computational support from the Princeton Center for Theoretical Science and the College of Arts and Sciences at Syracuse University.


\section*{Supplemental Materials and Methods}
\begin{figure}[h!]
\centering \includegraphics[width=0.3\textwidth]{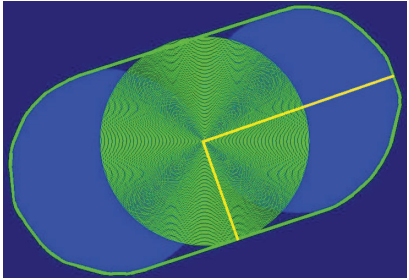}
\caption{\label{hull} The aspect ratio of the fusion aggregates was determined by finding the major and minor axis of the convex hull as indicated by the yellow lines.}
\end{figure}

\textbf{Aggregate fusion assay} For both experiments and simulations, the minor and major axis of the fusing aggregates were determined based on finding the convex hull in MATLAB (MathWorks, Natick, MA). The major axis corresponds to the maximum distance between points on the hull. The minor axis was determined by drawing circles (spheres in 3D) of increasing radii with center at the center of mass of the fusing aggregates, and determining the intersection with the convex hull (Fig.S~\ref{hull} and Supplemental movies).

\textbf{Mitotic index} We determined the mitotic index on fixed aggregates labeled with DAPI (Invitrogen), similar to the method described in Ref.~\cite{Kanki1997}. Aggregates were fixed in 4$\%$ paraformaldehyde-PBS and stored in 100$\%$ methanol at -20$\,^{\circ}\mathrm{C}$ until use. They were then rehydrated and stained with DAPI for 30~min at room temperature at 1:1000 in PBSTx (PBS containing 0.3$\%$ Triton x100) in the dark. Stained aggregates were mounted on tunnel slides and imaged on a Olympus DSU microscope (Tokyo, Japan). Mitotic figures and total nuclei counts were recorded in different regions of the aggregates and the total mitotic index MI (MI$=$ mitotic figure/total nuclei count) of 3-5$\%$ was obtained by averaging over multiple regions and aggregates.

\textbf{Tensiometer calibration}
The tissue surface tensiometer (TST) was calibrated by compressing a droplet of water of size comparable to cell aggregates (r$\sim$185$\mu$m) in commercial mineral oil (CVS pharmacy) and comparing the results with pendant drop measurements. For the pendant drop, we used the same chamber and microscope, with the lower compression plate retracted and a water filled syringe replacing the upper compression plate and balance. We analyzed two water drops while growing in size, with 8 and 7 measurements taken, respectively, using the analysis in Ref.~\cite{Anastasiadis1986}. All image and data analysis was done in MATLAB. The interfacial tension of oil and water was 18.8$\pm$0.4~dyne/cm for the TST ($n=4$ compressions) and 18.8$\pm$0.5~dyne/cm for the pendant drop. This falls within the range of previously published values~\cite{Norotte2008, Mgharbel2009}.

\subsection*{Mechanical model}
 We develop a model for zebrafish embryonic explants because they have many simplifying features, such as the absence of an extracellular matrix, rare cell divisions (SI and ~\cite{Kimmel}), and easy molecular manipulation~\cite{EMSthesis}. Additionally, these explants are cohesive and exhibit liquid-like behaviors such as cell sorting, ``rounding up'', and tissue fusion~\cite{EMSthesis, Schoetz2008, Krieg2008}.  The simplicity of the system allows us to study the fundamental mechanics of groups of cells in the absence of mitotic divisions and interactions with external boundaries.

Ideally, we would have liked to use the cell tracking data to calibrate an {\em existing model} for the mechanical behavior of tissues.  However, we discovered that models in the literature were either complex and thus difficult to constrain experimentally, or insufficient to explain our experimental data. For example, popular models for tissues such as Cellular Potts Models~\cite{Mombach, Graner1992, Shirinifard2009}, finite element models~\cite{chen2007, Brodland2009} and the Subcellular Element Model~\cite{sandersius_modeling_2008, jamali2010, sandersius2011} describe a single cell using tens to thousands of lattice points or elements, and a tissue is then composed of thousands of such cells. This level of detail can sometimes be necessary since individual cells are active and viscoelastic, which means that their equilibrium shape changes over long timescales in a history-dependent way.  If, on the other hand, one is only interested in the tissue behavior as a whole, it is not clear whether this computationally intensive and parameter-rich approach is necessary. Furthermore, while these cellular models accurately reproduce the experimental data, we would prefer a coarse-grained model for tissues that contains a few experimentally-accessible parameters and allows us to develop a physical intuition for the experimental system.

Several of these coarse-grained collective cell models exist but have limitations. Some models are intrinsically solid-like: either the tissue has a crystalline structure~\cite{Palsson} or the tissue flows only when there are significant rates of cell division or apoptosis~\cite{Farhadifar, Ranft, Staple}. A different class of ``active matter'' models is inherently fluid-like; these models exhibit macroscopic flow patterns even in the absence of cell divisions, but do not conserve momentum or include realistic cell interactions~\cite{Vicsek1995, Chate, Henkes}.  Since zebrafish explants exhibit fluid-like behavior despite very infrequent cell divisions, and cell interactions inside the explant must conserve momentum because cells do not interact with an underlying substrate, the models above cannot accurately capture this experimental system.

Therefore we develop our own model. We identify four general classes of interactions that occur between cells: resistance to shape changes, adhesion, damping, and active cell motility.

\textbf{Resistance to shape changes} The cytoskeleton and cell membrane resist changes to their shape and therefore individual cells resist being pushed into one another. Individual cells are viscoelastic and behave like liquids on long times and solids on short timescales. Although cells are polyhedral inside the tissue, they do have a characteristic radius and become spherical when alone in suspension.  For small deformations, the mechanical interaction between two spherically-symmetric elastic solids is hertzian~\cite{Landau_elastic}, while for two liquid membranes it is harmonic~\cite{Durian}, and a hybrid model was used for cells by Sato et al.~\cite{sato}.
\par
\textbf{Adhesion} Adhesive interactions between cells mediated by cadherins and other molecules generate additional forces. Although these interactions also have complex dynamics, we assume that the adhesive energy is proportional to the area of contact between the two cells.  Several contact mechanics models exist for interactions between elastic spheres that have a constant adhesive energy per unit contact area, which differ in their description of the stress concentration induced by the adhesive contact. One of theses models is the Deriaguin-Muller-Toporov (DMT) contact mechanics model~\cite{DMT, DMT2}, which works well for stiff spheres with a small adhesive energy, while the Johnson-Kendall-Roberts (JKR) potential~\cite{JKR} best represents soft spheres with large adhesive energies. We plan to investigate the JKR model in future work, but we chose the DMT model for this work for two reasons:

 First, the load $F^{int}$ is a simple analytic function of the overlap $\delta$.  This is not the case for the JKR model.  In addition, the adhesive energy term in the DMT model is exactly proportional to the ``buried surface area'' of two spheres. The buried surface area $\Delta SA$ is given by:
\begin{equation}
\Delta SA = 4 \pi R^2 \int_0^{\theta} \sin \theta d \theta = 4 \pi R^2 \left(1 - \cos \theta \right) = 4 \pi R \delta .
\end{equation}
If the total energy per unit area of contact is given by $\gamma$, then the adhesive force on each cell is $F = d/ d \delta (\Delta SA \gamma/ 2) = 2 \pi \gamma R$. Therefore the DMT model explicitly accounts for the adhesion being proportional to the area in contact; for similar reasons a linear adhesive potential also describes the adhesive energy between single molecules in solution~\cite{LiWingreen}.

\begin{figure}[h!]
\centering \includegraphics[width=0.3\textwidth]{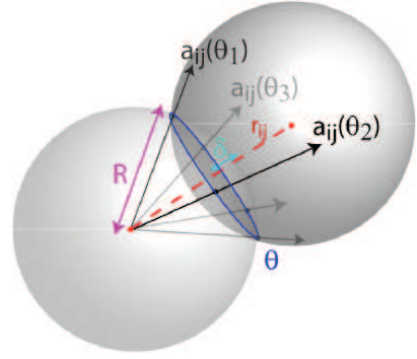}
\caption{\label{overlap} Schematic of overlapping spheres with radius $R$ and distance $r_{ij}$ between their centers. The overlap $\delta$ is shown in cyan. Protrusions effectively make new tensile contacts in a small region of overlap as indicated by the blue ring. Therefore active forces are directed along a family of vectors $a_{ij}$, parameterized by $\theta$, which extend from the center of each sphere to the overlap ring.}
\end{figure}

The DMT interaction potential approaches zero with a finite slope, i.e. two interacting spheres ``snap to contact'': at zero overlap the spheres experience a strong force inward for both loading and unloading.  As stated here, the model has no hysteresis because it does not include any energy dissipation. If some fraction $\alpha$ of the elastic energy is dissipated, however, the force experienced upon loading would decrease by a factor of $1/(1 + \alpha)$, while the force experienced upon unloading would increase by $1/(1 - \alpha)$~\cite{Zheng_adhesion}. For simplicity we initially set $\alpha =1$ (no dissipation), but understanding the energy balance between active forcing and dissipation is an interesting area of future research. Based on a  {\em harmonic} version~\cite{Durian} of the DMT contact mechanics model, the adhesive and repulsive interactions $\vec{F}^{int}_{ij}$ between two cells labeled $i$ and $j$ at positions $r_{i}$ and $r_{j}$ is given by:

\begin{equation}
\label{DMTeq}
\vec{F}^{int}_{ij} = \left( K \delta_{ij} - 2 \pi \gamma R\right)\widehat{r}_{ij}; \quad \vec{F}^{int}_{i} = \sum_{<ij>} \vec{F}^{int}_{ij},
\end{equation}
where $\delta_{ij}$ is the overlap distance $R_{i} + R_{j} - |r_{i}-r_{j}|$, and $\widehat{r}_{ij}$ is the unit vector in the direction of overlap.  $K$ is an effective spring constant, which models the cortical tension of a single cell, and $\gamma$ is an effective adhesive energy, which models a constant density of cadherins or other cell-cell adhesion molecules per unit cell surface area.

\par
\textbf{Damping} A third set of important forces are drag forces that dissipate energy. These could be generated by viscous liquids in the cellular environment or by frictional/adhesive forces from rubbing against adjoining cells. In the limit that there are no macroscopic shear flows in the tissue both can be modeled as a drag coefficient $b$ times the cell velocity~\cite{Durian}: $\vec{F}^{damp}_{i} = b v_{i}$. 

The natural time scale for the simulation, $\tau$ is given by
\begin{equation}
\tau = \frac{b}{K},
\end{equation}

where $b$ is a drag coefficient, and $K$ is a spring constant associated with the effective repulsive interaction between cells due to the cortical tension.  We estimate $R \sim 8 \mu m$, and the spring constant $K$ is roughly the elastic modulus of a single cell $Y  \sim 50 Pa$ (assuming that the Youngs modulus for a single cell is equivalent to the Youngs modulus for the entire aggregate because the tissue is confluent, see main text) times the cell radius.  The most difficult parameter to estimate is the drag coefficient. To get at least an order of magnitude estimate, we set $b = \Gamma R^3$, where $\Gamma$ is the intensive drag per unit volume. A very rough approximation for this number is the drag an object experiences when being pulled through a dense actin network; based on the work of Palmer et al~\cite{Palmer} we estimate this number as $\Gamma = \eta/ l_{mesh}^2$, with $\eta$ = the viscosity = $4 \times 10^{-3} Pa \cdot s$. The linear distance between actin filaments is estimated as $l_{mesh} = 100 nm$.  Therefore $b \sim 400 Pa \cdot s \cdot \mu m$, and an order of magnitude estimate for the natural timescale is $\tau \sim 1-10 s$.

The current model describes a ``mean-field'' viscous damping between a cell and its surroundings, but when all the cells are migrating together there is an important local contribution generated by friction between a cell and its neighbors. In addition, there could be some dissipation during the hysteritic ``snap to contact'' seen in all models of contact mechanics between adhesive objects. It would be interesting to investigate how these different mechanisms for dissipation balance the energy generated by active forcing and shape changes to generate a steady state.

\par
\textbf{Active forcing} Thermal fluctuations are not large enough to generate diffusion in cell aggregates, thus active biological processes must generate forces that enable cells to diffuse.  This is an involved process which has been modeled extensively in single cells -- individual cells change their shape, develop new contacts, exert tension on those contacts and contract to release old contacts.   Here, we model these active processes as an active force that changes the direction of the center of mass of each cell, which we denote $\vec{F}^{a}_{i} $.

The simplest model for active forcing is a random white noise process with amplitude $T$~\cite{Palsson,Palsson2}, but this does not capture some features of embryonic cells.  First, active forces are correlated in time: cells move in a directed fashion over a characteristic time required to disassemble and reassemble the actin network necessary for motility. We incorporate this in the model by requiring a cell to exert the same active force for a persistence time $p_t$, which is a model parameter.

Second, cells in confluent tissues move by exerting forces on neighbors, which means that active forces are spatially correlated, too~\cite{Ranft}. Because actin networks are contractile, adherent cells pull towards neighboring contacts (rather than pushing away). Our model captures this inherent asymmetry in force generation by including a special spatial structure for the ``active noise'' .   Within our model, contact between cells $i$ and $j$ is made along a ring of vectors $\widehat{a}_{ij}(\theta)$ which can be expressed simply in terms of the particle centers and radii ( SI Fig. S2). 

The model assumes that at each timestep two cells exert equal and opposite forces on each other at a randomly chosen point along that ring of contact. The magnitude of the force $\boldsymbol{\sigma}$ taken from a chi distribution with $k= 3$ degrees of freedom and variance $\sigma^2$, which is the simplest assumption for the random magnitude of a three-component vector. Therefore $\vec{F}^{a}_{ij} = \boldsymbol{\sigma} \; \widehat{a}_{ij}$.

\subsection{Simulation Initialization}
Unless otherwise noted, all simulations were for 1000 cells.  The total volume of each cell was normalized to unity, with the radii chosen from a triangle distribution with mean $R = (3/(4 \pi))^{1/3} \sim 0.062$ and width $0.1 R$. At each time step we integrate the set of equations given by
\begin{equation}
\label{dim_model}
\frac{\widetilde{dr_{i}}}{d\widetilde{t}} = - \sum_{j} \left[\left( \widetilde{\delta}_{ij}  - \widetilde{\Gamma} \right) \widehat{r}_{ij} + \widetilde{\sigma} \; \xi \; \widehat{a}_{ij} \right],
\end{equation}
using a standard implicit integration method~\cite{Durian} with timestep $dt= 0.01 \tau$. The results did not depend on the timestep up to $dt= 0.1 \tau$.   The cell centers of mass (COM) were initialized from a uniform random point pattern in a three-dimensional periodic box of length unity using a conjugate gradient routine and the interaction potential given by Eq.1 in the main text alone, generating a cube of tissue with packing fraction unity.  Because most experiments are performed on initially spherical tissues, we then evolved the cell COMs according to Eq.~\ref{dim_model} with a relatively large active force magnitude $\widetilde{\sigma} = 1$ and persistence time $\widetilde{p_t} = 0.1$ without periodic boundary conditions so that the group of cells would round up into a ``droplet''.  This droplet COM configuration was then used to initialize the remainder of the simulations.

  For the simulations of tissue compression, we represent the compression plates by a dense packing of wall particles that interact with the simulated cells. To ensure that the cells in the tissue can not push through the wall particles, a cell labeled $i$ interacts with a wall particle labeled $j$ via a hard-core repulsive term proportional to $\delta_{ij}^{-3} -1$, where $\delta_{ij}$ is the overlap of the radii of the two objects as described above.  The wall particles are very tightly packed, at packing fraction 1.  We initialize the $z$-position of the wall particles so they are not touching the droplet, and as we integrate the active model equations for the cells (Eq.~\ref{dim_model}), we move the top wall downward with a velocity $v_{wall} = $ cell radii/$\tau$, until the aggregate is compressed in the $z$-direction by 10\% of its initial $z$-radius. At that point the wall stops moving, and the cells continue to evolve according to the model equations for 500 natural time units.  At each time step we record the net force of the cells on the wall, which is the sum of all of the cell-wall interaction forces during a given time step.

\subsection{Surface tension parameter extraction}
\begin{figure}[h!]
\centering \includegraphics[width=0.45\textwidth]{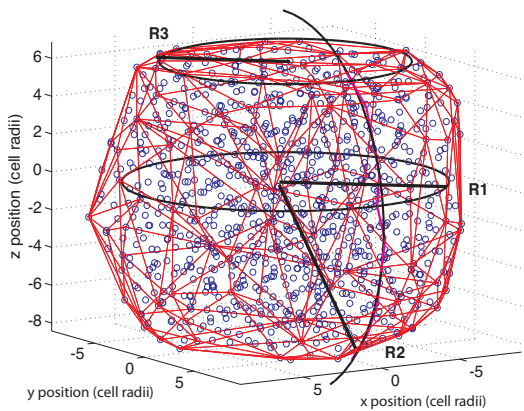}
\caption{\label{TSTgeom} Sample final state for a TST simulation, indicating the geometry of the simulated cells and extracted curvatures. }
\end{figure}

To calculate the surface tension of the experimental aggregates and water drops, we used Eq. 4 in the main text, because as pointed out by Ref.~\cite{Mgharbel2009}, it is difficult to accurately measure the radius of the tissue in contact with the upper plate, $R_3$. For the simulated aggregates, we have direct access to information about which cells are in contact with the upper plate in TST simulations which we used to calculate $R_3$.  Fits to this variable are more robust than those for $R_1$ and $R_2$, because the simulated aggregates are small and therefore have larger fluctuations in those variables as compared to the experimental aggregates. For the simulated aggregates we therefore calculated the surface tension $\gamma$ directly using Laplace's law~\cite{EMSthesis}:
\begin{equation}
\label{gamma_TST}
\frac{F_{eq}}{\pi R_3^2} = \left( \frac{1}{R_2} + \frac{1}{R_1} \right) \gamma.
\end{equation}
$F_{eq}$ is the steady-state force on the tensiometer upper plate, and we extract $R_1$, $R_2$, and $R_3$ by fitting circles to the isotropically average centers of mass of cells in the simulations, as shown in Fig.~\ref{TSTgeom}

  As discussed by Ref.~\cite{Mgharbel2009}, there is an additional contribution to the surface tension for small fluid droplets that is proportional to the perimeter of contact $2 \pi R_3$ between the aggregate and the plate, and it also depends on the contact angle $\theta$.  In using equation~\ref{gamma_TST}  to calculate the surface tension, we have implicitly assumed that the contact angle is small.  While this holds for the experimental aggregates as the compression plates were specifically treated to minimize aggregate adhesion (indicating that Eq. 4 in the main text is approximately equal to SI Eq.~\ref{gamma_TST} for those tissues), it is not clear in the case of simulated aggregates.  We attempted to fit for the contact angle in our simulations, but the fits were not robust and contributed wildly varying terms to the final estimate of the surface tension.  In contrast, the values for the surface tension that we found by using SI Eq.~\ref{gamma_TST}  and assuming a vanishing contact angle were robust, which led us to use this approach for the simulation values quoted in this manuscript.  In the future, it would be interesting to systematically study the line tension and surface tension in simulated droplets of different sizes to understand if a thermodynamic expression like the one used by Ref.~\cite{Mgharbel2009} is in fact appropriate for our model, or if there are systematic deviations.  This would also be interesting to study in very small experimental tissues.

\section*{Supplemental Results}

\subsection{Calibrating the model based on nuclei tracking data}

To calculate the scaling exponent for the mean squared displacement (MSD), $\alpha$, we find the maximum value of the derivative $d log (MSD)/ d log (t)$ and take the average of that derivative over a window of data above the crossover timescale $t_c$.  This window is 10 \% of the data points with squared displacements which ranked between the 80 \% and 90 \% largest. In addition, we do not calculate the MSD for the first 20 natural time units of the rounded droplet simulation to ensure that we avoid transient effects from the system initialization. $D$ is calculated as $(1/6) d (MSD)/dt$ averaged over the same window as $\alpha$.

\begin{figure}[h!]
\centering \includegraphics[width=0.5\textwidth]{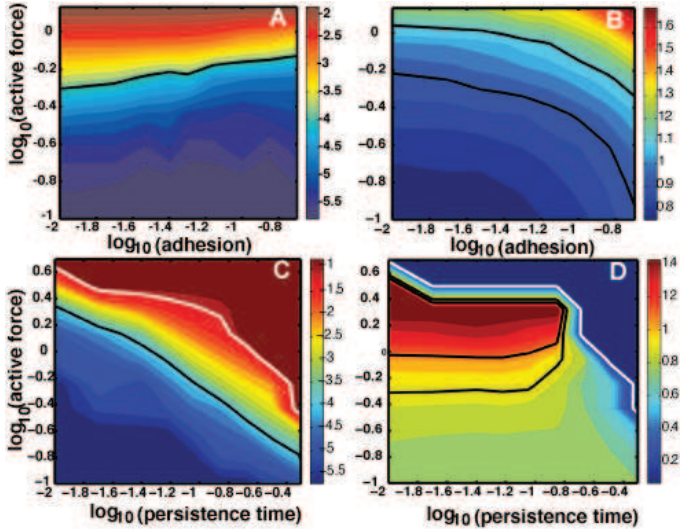}
\caption{\label{SIphase} (A, B) Phase diagrams as a function of the active forcing magnitude and adhesion with $p_t = 0.10$. Regions in between black lines are consistent with experimental results. (C, D) Phase diagrams as a function of the active forcing magnitude and persistence time  with $\Gamma = 0.04$. Regions between black lines are consistent with experimental results. Regions above the white lines are numerically unstable and are not consistent with experimental results. For (A) and (C) , colorscale is log $_{10}$ of the diffusion constant $D$, while for (B) and (D), colorscale indicates the packing fraction. }
\end{figure}

In our model, the effective radius determines the volume of each cell.  To ensure that the model is consistent with no extracellular gaps in the tissue, we require that the sum of all the single cell volumes, determined from their effective radii, matches the total volume of space taken up by the group of cells as determined from the average radius of the entire aggregate during a simulation. Therefore, we calculate the average volume ratio (the inverse packing fraction) for a simulated aggregate as a function of two of the model parameters (magnitude of the active force $\widetilde{\sigma}$ and ratio between the adhesive energy and cortical tension $\widetilde{\Gamma}$), with the persistence time fixed at $0.1$.  The uncertainty in the estimated outer radius of the simulated aggregate is approximately one cell radius; this generates an uncertainty of about 10 \% in the volume ratio.  SI Figures~\ref{SIphase} (B) and (D) show the packing fraction phase diagrams for our simulations.  The lower, curved hatched region in Fig 3(A) in the main text denotes the region of the parameter space where the packing fraction is within 10 \% of unity. Since the packing fraction is not a strong function of adhesion for values of $\widetilde{\Gamma} < 0.05$, we fix $\widetilde{\Gamma} = 0.04$ and study the model as a function of the active force magnitude $\widetilde{\sigma}$ and persistence time $\widetilde{T}$.  The lower cross-hatched region in Fig~3(B) in the main text denote simulations with packing fraction close to unity.

We also study the single-cell dynamics within the framework of this model; since the simulation has no rotation or drift we analyze the average MSD as a function of time.  For each simulation we calculate the scaling of the power law relationship between the MSD and time $\alpha$,  the diffusion constant $D$, the sub-diffusive to diffusive crossover timescale $t_c$ , and the dimensionless product $\tau^* = D t_c /R^2$. It is difficult to calculate $\alpha$ when the total displacements are small (i.e less than a cell radius), and we find that in our simulations $\alpha$ is close to unity if and only if $D > 1 \times 10^{-4}$ in natural units.  Therefore we use $D$ (which is less noisy) as a proxy for the scaling exponent $\alpha$, and investigate how $D$ varies with our three dimensionless model parameters.  The phase diagrams for the diffusion constant are shown in SI Fig.~\ref{SIphase} (A) and (C), and the black lines indicate the regions consistent with experiments. The upper hatched region in Fig~3 (A) in the main text is where $D > 1 \times 10^{-4}$. Once again we find that $D$ does not vary strongly with $\widetilde{\Gamma}$ and we therefore study $D$ as a function of active force and persistence time at fixed $\widetilde{\Gamma} = 0.04$. The region with slanted lines in Fig~3(B) in the main text is where $D > 1 \times 10^{-4}$. Solid regions in both of these plots indicate regions where cells overlap significantly (in disagreement with experimental observations) and the simulations eventually become numerically unstable.

Combining our results for the packing fractions and $D$, we find a small region in our simulation parameter space that is consistent with the experimental data, denoted by the crosshatched areas in Fig~3 (A) and (B) in the main text. The dimensionless parameter $\tau^* = D t_c /R^2$ was found to be approximately $0.07$ for experimental ectoderm explants,  and therefore we restrict simulation parameters to regions where $0.03 < \tau^*  < 0.08$.  Because $t_c$ and $D$ are quickly-varying observables, the results are not sensitive to the exact cutoffs chosen (see the sensitivity analysis below). There is a very small region of parameter space that satisfies the three constraints on $D$, $\tau$, and the packing fraction based on those values for the experimental tissues, and it consists of two points denoted by red and green symbols in Fig.~3 (A,B) in the main text.

   The best-fit point in phase space for ectoderm, denoted by the red circles in Fig.~3 (A,B) in the main text, has a diffusion constant  $D = 2.1 \times 10^{-4}$ cell radii$^2$ per unit time, where the time is in natural units for the simulation $b/(K)$.  This natural unit is roughly the time it takes for a cell to relax to mechanical equilibrium once it has moved past another cell.   Although it is difficult to determine directly because the damping coefficient $b$ is not well-constrained experimentally, we simply equate the simulation diffusion constant with the experimentally determined diffusion constant for the ectoderm, $D = 0.22 \mu m^2 /$min and conclude that with the cell radius $R = 8 \mu m$, the natural simulation timescale unit is $4$ seconds.

     To estimate the mesendoderm parameters, we assume that the natural timescale for the mesendoderm simulation is the same as that for the ectoderm simulation, and use the experimentally observed mesendoderm diffusion constant $D=0.60 \mu m^2/$ min to constrain the model parameters. Under this approximation, we find that $\widetilde{\sigma} = 0.88$, $\widetilde{\Gamma} \sim 0.04$, and $\widetilde{p}_t = 0.08$ are the best-fit parameters for the mesendoderm tissue, as denoted by the green cross in Fig.~3 (A,B) in the main text. Table~1 in the main text summarizes the best fit parameters and conversion factors from our simulation model for both tissue types.

To determine how sensitive the observables $O$ (i.e. the diffusion constant $D$ and packing fraction $\phi$) are to changes in model parameters $x_i = \{ \widetilde{\Gamma}, \widetilde{\sigma},\widetilde{p_t}\}$, we calculate sensitivity parameters $(\partial O/ \partial x_i)|_{x0i}*(x_{0i}/O)$ evaluated at the best-fit parameters $x_{0i}$ for ectoderm shown in Table~1 in the main text. 

\begin{table}[h!]
\begin{tabular}{  l | c |c  }


 & $D$ & $\phi$ \\
 \hline
  $\widetilde{\Gamma}$ &  0.95 & 0.04 \\
  $\widetilde{\sigma}$ & 6.6 & 0.36 \\
  $\widetilde{p_t}$ & 2.07 & -0.17 \\
\end{tabular}
\caption{\label{table_best} Sensitivity parameters $(\partial O/\partial x_i)|_{x0i}*(x_{0i}/O)$.}
\end{table}

This demonstrates the the diffusion constant is much more sensitive to changes in model parameters than the packing fraction, and that both observables are most sensitive to the magnitude of active forcing $\widetilde{\sigma}$ and not very sensitive to changes in adhesion $\widetilde{\Gamma}$.

\subsection{Tissue structure and dynamics}

\begin{figure}[h!]
\centering \includegraphics[width=0.5\textwidth]{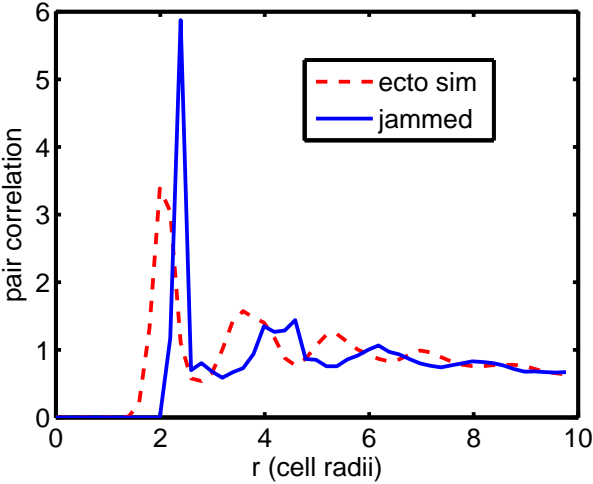}
\caption{\label{gofr} Pair correlation function for simulated tissue. The red dashed line corresponds to the best-fit ectoderm simulation, which becomes diffusive at long timescales. The blue solid line corresponds to simulated tissues with $g = 0.04$, $\sigma = 0.57$ and $p_t = 0.01$ that do not become diffusive.  The large first peak and split second peak seen in the blue curve are signatures of jamming and icosahedral order.}
\end{figure}

In contrast to other models for active tissues~\cite{Palsson, Palsson2}, our cells remain disordered even when the system has well-defined self-generated boundaries.  When the model is simulated in a parameter regime that is highly diffusive, the pair correlation function for nuclei centers is indistinguishable from that for a liquid (SI Fig.~\ref{gofr} dashed line). The correlation function has decaying peaks that correspond to coordination shells.  The pair correlation functions for the best-fit parameters for the ectoderm and mesendoderm tissues are similarly liquid-like.  This model has an apparent jamming or glass transition, corresponding to the region where $D < 10^{-4}$ in the phase diagrams, however, and there the pair correlation function shows typical characteristics of jamming, such as a higher peak in the first coordination shell and a split second peak generated by icosahedral order (SI Fig.~\ref{gofr} solid line).  The limited statistics available in the experimental system prevent us from being able to determine whether the pair correlation function is liquid-like or glassy, but it is clear from confocal images of two-dimensional slices that the cell packing is also disordered.

\begin{figure}[h!]
\centering \includegraphics[width=0.3\textwidth]{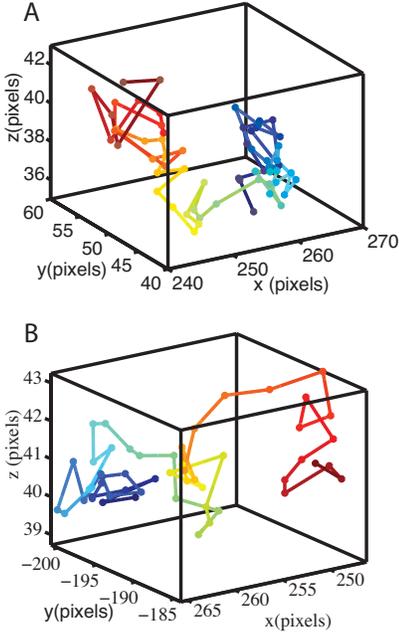}
\caption{\label{caging} Two example trajectories suggestive of cage-breaking events from the experimental ectoderm nuclei tracking data.}
\end{figure}

Because the ectoderm and mesendoderm tissues are apparently in the liquid phase close to the glass/jamming transition on our phase diagram, we anticipate that the experimental tissues would exhibit some signatures of a ``supercooled'', near-glassy state.  One signature of such a state is a transition from sub-diffusive to diffusive behavior, which we do see in the experimental MSD data shown in the main text.  If the origin of this behavior is similar to that in a supercooled liquid, we expect to see cells move around a small amount inside ``cages'' generated by their neighbors, and then ``jump'' a distance roughly corresponding to a cell diameter due to a relatively rare cage-breaking' event.  We viewed the trajectories of hundreds of cells, and saw that quite a few exhibited this type of behavior.  SI Fig.~\ref{caging} (A, B) show two sample trajectories from nuclei in an ectoderm explant that appear to have cage-breaking events.  We also studied this effect systematically by analyzing the non-gaussian parameter as discussed in the main text.

\section*{Supplemental Movies}
\textbf{Movie S1} Rounding-up of zebrafish ectoderm tissue, acquired at 1 frame every 2 min using brightfield imaging. Displayed at 20 frames/sec.\\
\textbf{Movie S2} Zebrafish ectoderm tissue fusion, acquired at 1 frame every 2 min using brightfield imaging. Displayed at 20 frames/sec.\\
\textbf{Movie S3} Dynamics of convex hull analysis of data from movie S2. Displayed at 20 frames/sec.\\
\textbf{Movie S4} Simulation of fusion of aggregates. Blue dots are cell centers, red lines illustrate the convex hull. \\



\bibliographystyle{rsc}
\bibliography{tissueall_v07}


\end{document}